\documentclass[fleqn,twoside]{article}
\usepackage{espcrc2}
\usepackage[dvips]{graphicx}
\usepackage{epsf}
\usepackage{amssymb}
\usepackage{bm}

\newcommand{\AmS}{{\protect\the\textfont2
  A\kern-.1667em\lower.5ex\hbox{M}\kern-.125emS}}

\newcommand{\beq}{\begin{equation}}
\newcommand{\eeq}{\end{equation}}

\newcommand{\nn}{\nonumber \\ }
\def\abs#1{ \left| #1 \right| }

\title{
{\Large Orthopositronium Decay Spectrum using NRQED }
}

\author{
       Pedro D. Ruiz-Femen\'\i a
\address{
       {\em Max-Planck-Institut f\"ur Physik (Werner-Heisenberg-Institut),}\\
       {\em F\"ohringer Ring 6, 80805 M\"unchen, Germany}}
}

\begin{document}

\begin{abstract}
As noticed in Ref.~\cite{pestieau}, the Ore-Powell's
classical calculation of the o-Ps$\to 3\gamma$ decay amplitude 
does not fulfill Low's theorem requirements for the low energy end of the
photon spectrum. We reanalyze the implications of
Low's theorem applied to this decay considering the interplay 
between the soft photon limit and the energy scales
present in the $e^+e^-$ system. For energetic photons, the spectrum agrees with the
Ore-Powell result, but deviates from it when the photon energy is comparable to the
positronium binding energy. In this region it is found that 
bound states effects are essential to reach agreement with Low's 
theorem and can be accounted for in the framework of non-relativistic
QED. 
\end{abstract}
\maketitle


\section{Introduction}

The spin-triplet (orthopositronium: o-Ps) $3\gamma$ annihilation rate
was first obtained by 
Ore and Powell~\cite{OrePowell} in the late forties and can be found 
in many textbooks. Although the leptons inside the positronium atom
are bound by the Coulomb potential, the effects of such binding have been
commonly neglected in the calculation of the o-Ps decay rate, being
the process dominated by the short distance part where the
leptons annihilate into three photons.  
The Coulomb force will be expected to affect the annihilation process
appreciably only when a photon of small momentum is involved, as already
quoted in the original paper of Ore and Powell. The analysis of the
low energy photon spectrum of the $\mathrm{oPs}\to 3\gamma$ decay made 
in Ref.~\cite{our} 
takes into account these novel features by the use of the 
non-relativistic effective field theory methods 
developed to study bound states
in QED and QCD~\cite{caswell,bbl,lmr,ps}.

The later work was motivated by the observation of Pestieau and Smith~\cite{pestieau},
that the Ore-Powell calculation of the o-Ps decay is in apparent
contradiction with Low's theorem. While Low's theorem applied
to this decay requires that the spectrum vanishes as $E_\gamma^3$, with $E_\gamma$ the
energy of the radiated photon, the standard calculation predicts a ${\cal
O}(E_\gamma)$ behavior for $E_\gamma \to 0$. We show 
that  the Ore-Powell calculation can be reconciled with Low's theorem if the latter is
carefully applied considering all the energy scales present in the photon decay 
spectrum: the electron mass $m$, the
binding energy of order $m \alpha^2$, and the hyperfine splitting between the singlet
and triplet states of order $m \alpha^4$. The Ore-Powell computation is valid for
photon energies $E_\gamma \gg m \alpha^2$. When $E_\gamma$ is of order $m\alpha^2$, the
Ps binding energy can not be neglected. The decay amplitude depends on a sum over an
infinite set of excited Ps states, and can be written in terms of a Ps structure
function which was computed in Ref.~\cite{our} using NRQED to leading order in
the velocity $(v)$ expansion. When $E_\gamma$ is of order $m \alpha^4$, the decay amplitude is
dominated by the o-Ps $\to$ p-Ps transition and it can be shown that the photon spectrum
crosses over from an $E_\gamma$ behavior above the structure function region to an
$E_\gamma^3$ behavior below the p-Ps resonance region.

\vskip -.5cm
\section{Low's theorem and o-Ps $\to\;3\gamma$ decay}\label{sec:Low}

Low's theorem~\cite{low} gives the amplitude for soft photon emission in the scattering
of charged particles. It states that the first two terms of the series expansion in
powers of the photon energy of a radiative amplitude $X \to Y \gamma$ may be obtained 
from a knowledge of the corresponding nonradiative amplitude $X \to Y$:
\begin{equation}
\epsilon_{\mu}{\cal M}^{\mu}=\frac{{\cal M}_0}{k}+{\cal M}_1+{\cal O}(k).
\label{2.1}
\end{equation}
One finds that ${\cal M}_0$ and ${\cal
M}_1$ are independent of $k$ and completely determined from the nonradiative amplitude
$T_0$, its derivatives in physically allowed kinematic directions, and the
electromagnetic properties of the particles involved~\cite{low}.

The term ${\cal M}_0$ arises from the emission of a photon by ingoing or outgoing
charged particles and is proportional to $T_0$ times the universal factor  $-Q_i\,
\epsilon\cdot p_i/k\cdot p_i$, summed for all the external lines in the diagram.  Note
that if the nonradiative process involves no moving charged particles or is forbidden
due to some selection rule, then ${\cal M}_0$ is identically zero. The term ${\cal
M}_1$ can be expressed as a function of the magnetic moments, the amplitude $T_0$ and
its derivatives with respect  to internal variables which are not subject to
constraint, e.g.\ the energy and angles.

Combining the amplitude behavior with that of the phase-space, the low-frequency form
of the photon spectrum is
\begin{equation}
 \frac{{\rm d}\Gamma}{{\rm d}E_\gamma}=\frac{A}{E_\gamma}+B+{\cal O}(E_\gamma),
 \label{2.2}
\end{equation}
where $A$ is proportional to $\abs{{\cal M}_0}^2$ and $B$ is the ${\cal M}_0 {\cal
M}_1$ interference term. If ${\cal M}_0$ vanishes, the soft photon decay spectrum is of
order $E_\gamma\,{\rm d}E_\gamma$; if both ${\cal M}_{0}$ and  ${\cal M}_{1}$vanish, it
is of order $E_\gamma^3\,{\rm d}E_\gamma$.

In the three-photon decay of o-Ps, one of the photons can have an  arbitrarily small
energy. The process can be then viewed as the radiative version of the o-Ps$\to
2\gamma$ decay. As the two-photon decay of o-Ps is not allowed by charge conjugation
invariance, the direct application of Low's theorem yields  ${\cal M}_{0,1}=0$ so that
the o-Ps$\to 3\gamma$ amplitude is of order ${\cal O}(E_\gamma)$, and the decay
spectrum is
\begin{equation}
\frac{{\rm d}\Gamma_{{\rm oPs}\to 3\gamma}}{{\rm d}E_\gamma}\sim E_\gamma^3
\label{2.3}
\end{equation} 
as $E_\gamma\to 0$. 
This is in contradiction with the Ore-Powell spectrum,
\begin{eqnarray}
{{\rm d}\Gamma_{3\gamma} \over {\rm d} x}  &=&  
{2 m \alpha^6 \over 9 \pi} \left[ {5 x \over 3} + \mathcal{O}(x^2) \right],
\ \ x={E_\gamma \over m}
\label{2.4}
\end{eqnarray}
which
vanishes linearly with $E_\gamma$, as pointed out 
in Ref.~\cite{pestieau}.

To understand the origin of the contradiction, it is worth noting that in the
derivation of Low's theorem, one takes the limit $E_\gamma \to 0$ and neglects all
states other than those degenerate with the incoming and outgoing states, i.e.\ one
uses $E_\gamma \ll \Delta E$, where $\Delta E$ is the energy gap to excited states. The
amplitudes ${\cal M}_{0,1}$ depend on the charge and magnetic moment couplings between
all the intermediate states degenerate with the initial or final states. A more general version of Low's
theorem gives the decay spectrum for small $E_\gamma$ without taking the strict
$E_\gamma \to 0$ limit. One treats all states with $\Delta E \ll E_\gamma$ as
degenerate states, and includes them in the computation of charge and magnetic moment
matrix elements for the purposes of Low's result. Which states are included in Low's
theorem then depends on the magnitude of $E_\gamma$.

In the case of Ps decay, consider the case where the photon energy is much larger than
the binding energy. Then all Ps states (including o-Ps, p-Ps, radial excitations, etc.)
are degenerate for the purposes of Low's theorem. In this case, there is a non-zero
magnetic dipole matrix element between o-Ps and p-Ps, so that ${\cal M}_1$ does not
vanish in this extended space of states. As a result, the decay spectrum vanishes
linearly with $E_\gamma$. This is the approximation under which the Ore-Powell
calculation is valid. For energies much smaller
than the o-Ps--p-Ps hyperfine splitting, p-Ps as well as radial excitations are treated
as excited states, the matrix element ${\cal M}_1$ vanishes, 
and the spectrum is of order $E_\gamma^3$.

The discussion above could be also applied to the
$3\gamma$ annihilation of excited Ps states, like the $^1P_1\to 3\gamma$ decay~\cite{Gromes}.

\vspace{-.1cm}
\section{Orthopositronium Decay Amplitude}\label{sec:ortho}

\begin{figure}
\begin{center}
\includegraphics[width=3cm]{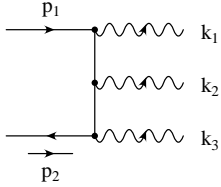}
\end{center}
\vspace{-1.3cm}
\caption{Three photon annihilation graph. The graph is summed over the
 $3!$ permutations of the photons.\label{fig:3gamma}}
 \vspace{-.4cm}
\end{figure}

The Ore-Powell annihilation amplitude can be obtained from the free-particle decay amplitude
given, to lowest order in $\alpha$, by the graph in 
Fig.~\ref{fig:3gamma}.
If the o-Ps momentum space wavefunction is $\phi_o(\mathbf{p})$ and the free $e^+e^-$
annihilation
amplitude is $A(\mathbf{p})$, the bound-state decay amplitude
to lowest order in $v$ is
\begin{eqnarray}
\int {{\rm d}^3 {\mathbf p} \over (2 \pi)^3} A(\mathbf{p}) \phi_o(\mathbf{p})
\simeq A(0) \psi_o(0) \ ,
\label{3.1}
\end{eqnarray}
where $\psi_o(0)$ is the o-Ps position space wavefunction 
at the origin. Spin averaging $\abs{A}^2$ and integrating over
the three-body phase-space yields the differential decay rate 
written in Eq.~(\ref{2.4}).

The non-relativistic effective theory 
computation (see details in Ref.~\cite{our})
provides a systematic way of including bound state effects 
in the o-Ps decay amplitude. The
NRQED Hamiltonian is constructed to reproduce the QED amplitude
when binding effects are neglected.
Once the NRQED Hamiltonian has been determined, it can be used to
compute the decay including binding corrections.

The bound state dynamics is described by the Coulomb Hamiltonian for an $e^+e^-$ system
in interaction with the quantized electromagnetic field:
\begin{eqnarray}
H &=& H_0+H_{\mathrm{int}} \nn
H_0 &=& \frac{\mathbf{p}^2}{m}-\frac{\alpha}{r} \nn 
H_{\mathrm{int}} &=& -{e\over 2m}\, [{\bm{\sigma}_{\phi}+\bm{\sigma}_{\chi}}]
\cdot{\mathbf{B}}
-e\,{\mathbf{x}}\cdot{\mathbf{E}}
\label{5.01}
\end{eqnarray}
with $\mathbf{x}$ and $\mathbf{p}$ the relative position and relative
momentum of the pair of
leptons, and
$\bm{\sigma}_{\phi},\,\bm{\sigma}_{\chi}$ the Pauli matrices acting on the electron and
positron spinors. The electromagnetic interactions are through
multipole interactions with the electric and magnetic fields and
only the electric and magnetic
dipole interactions are shown in Eq.~(\ref{5.01}), the higher multipoles being of
higher order in $v$.

The Coulomb Hamiltonian $H_0$ is the leading term in the velocity power counting. 
The kinetic energy and Coulomb potential are of the same order in $v$. The energies and
wavefunctions of $H_0$ are thus the Coulomb wavefunctions with reduced mass $m/2$.
The electric and magnetic dipole interaction terms are treated as perturbations. 
The o-Ps decay amplitude has both short distance and long-distance contributions.
The long distance part of the o-Ps decay amplitude entails the radiation
of the soft photon ($E_{\gamma}\ll m$) from the o-Ps bound state,
whereas the short distance part is given by the on-shell $e^+e^-\to 2\gamma$
QED decay amplitude where the photons are hard ($E\sim m$). 

\begin{figure}
\begin{center}
\includegraphics[width=3.3cm]{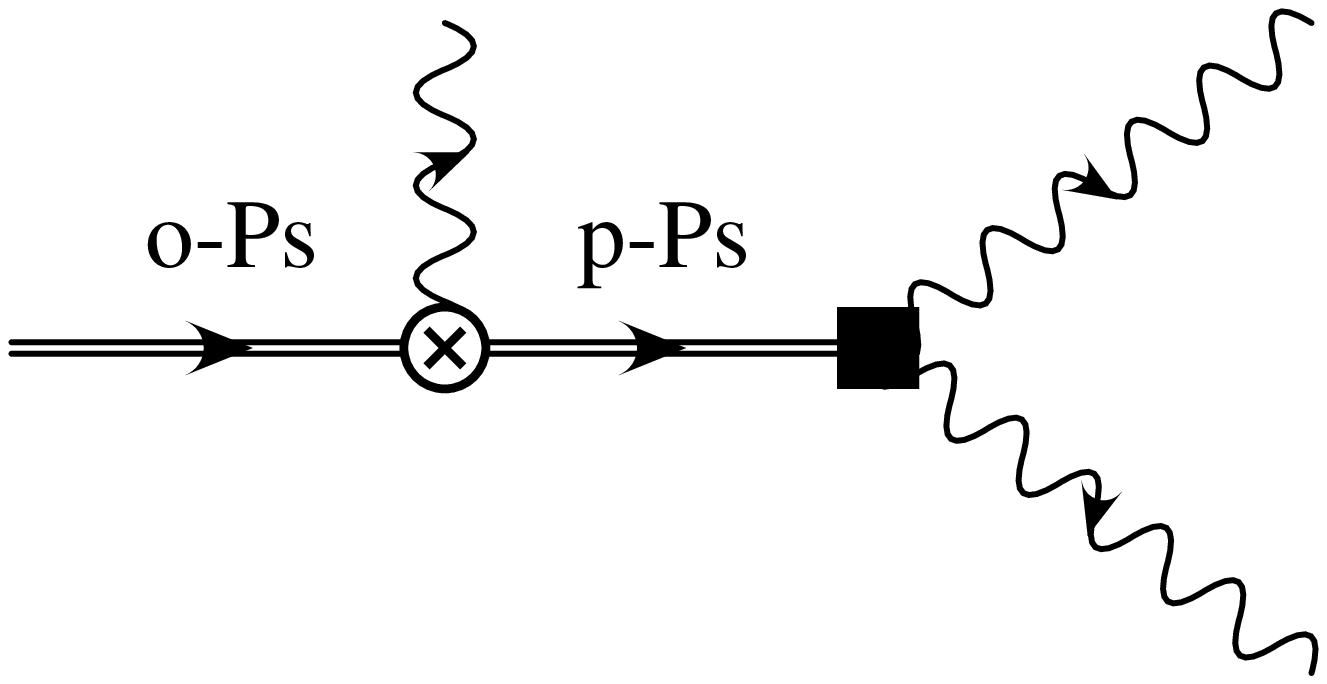}
\put(-58,2){$(a)$}
\put(47,2){$(b)$}
\vspace{-2cm}
\hspace{0.2cm}
\includegraphics[width=3.3cm]{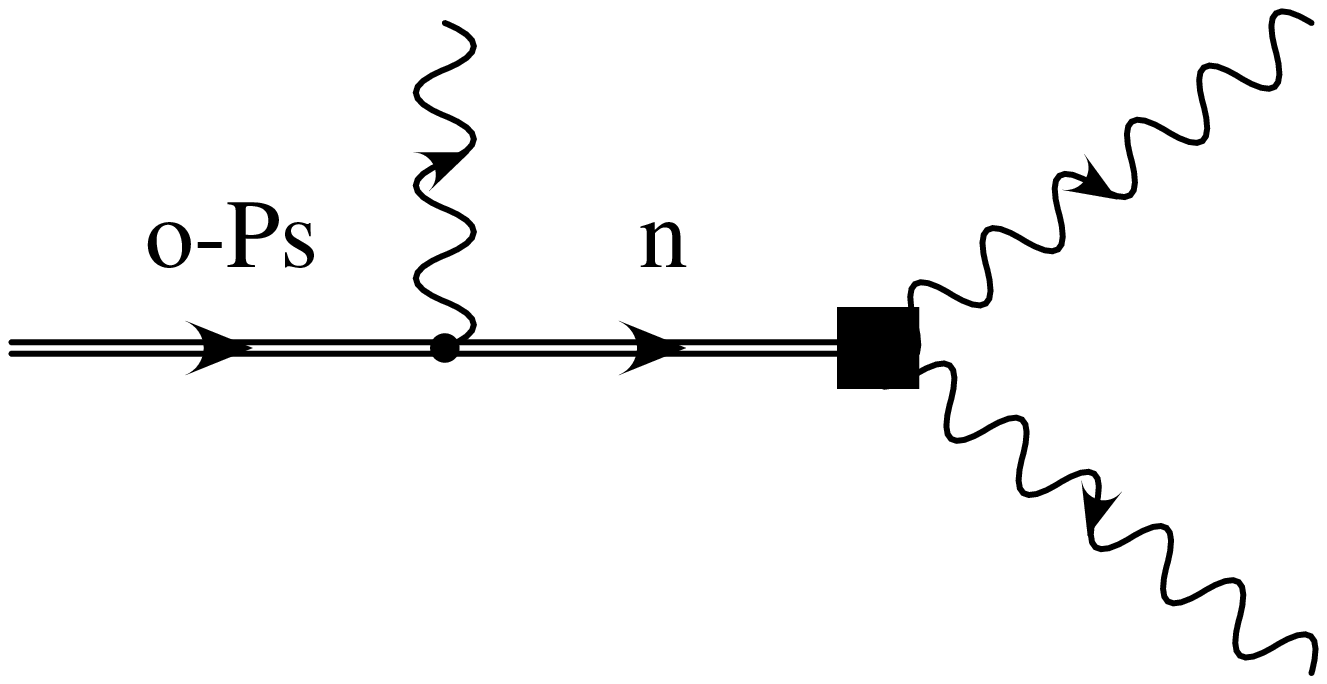}
\vspace{1.5cm}
\caption{$(a)$ Magnetic dipole graph for o-Ps annihilation. The solid square denotes the 
p-Ps annihilation vertex. $(b)$ Electric dipole transition for o-Ps decay. 
The solid square denotes the $n ^3P_{0,2}$ annihilation vertex\label{fig:eftgraphs}}
\end{center}
\vspace{-1cm}
\end{figure}

The effective theory graphs which describe the o-Ps$\to 3\gamma$ decay are shown in 
Fig.~\ref{fig:eftgraphs}.
The magnetic term in $H_{\mathrm{int}}$ can induce a $1\,^3S_1\to 1\,^1S_0$ transition
between Coulomb $e^+e^-$ states. The only allowed intermediate 
state in the dipole approximation is the p-Ps ground state, with energy $E_p$, 
so that $E_o-E_p=\Delta E_{\,\mathrm{hfs}}$.
The electric dipole term $-e\,{\mathbf{x}}\cdot{\mathbf{E}}$ in $H_{\mathrm{int}}$
can change orbital angular momentum by one unit, allowing for transitions from ground state
o-Ps to $n\,^3P_{0,2}$ states ($n\ne 1$).  
 
To compute the o-Ps decay amplitude matching coefficient, one compares the effective
theory result with the QED computation which neglects bound state effects.
For the matching computation, the energy separation between o-Ps and the intermediate
states, $E_o-E_n\sim {\cal O}(m\alpha^2)$ is taken to be much smaller than the photon
energy $E_\gamma$.

\vspace{-0.1cm}
\section{Results for the low energy spectrum}\label{sec:denomi}

The o-Ps$\to 3\gamma$ decay spectrum from our effective theory is written in terms of the 
magnetic and electric amplitudes obtained from the graphs in
Fig.~\ref{fig:eftgraphs}~\cite{our},
\begin{eqnarray}
{ {\rm d}\Gamma \over  {\rm d}x }  &=&{m \alpha^6 \over 9 \pi} x \left[\abs{a_m}^2 +  \frac
7 3 \abs{a_e}^2\right] .
\label{4.1}
\end{eqnarray}
For the magnetic term $a_m$, it is found 
\begin{eqnarray}
a_m &= & { E_\gamma\over E_\gamma-\Delta E_{\,\mathrm{hfs}}  - i \Gamma_p/2 }  ,
\label{4.2}
\end{eqnarray}
\begin{eqnarray}
\Delta E_{\,\mathrm{hfs}} = \frac{7}{12}\, m \,\alpha^4\ \ \ , \ \ \ 
\Gamma_p={1\over 4} m\alpha^5 .
\label{4.3}
\end{eqnarray}
The electric amplitude is written in terms of the $p$-wave Coulomb Green's
function~\cite{our},
which contains the sum over $n ^3P_{0,2}$ states in Fig.~\ref{fig:eftgraphs}b:
\begin{eqnarray}
a_e (E_\gamma) =  {4\pi E_\gamma\over \psi_o(0)} \int_0^{\infty} {\rm d}  y\,y^4\,
G_1(0,y;E_\gamma)\,\psi_o(y).
\label{4.4}
\end{eqnarray}
A closed analytical formula for this integral has been given in Ref.~\cite{Voloshin}.
The magnetic and electric amplitudes behave as
\begin{eqnarray}
a_m(E_\gamma) &=& \left\{ \begin{array}{cl}
1 &\qquad E_\gamma \gg  \Delta E_{\,\mathrm{hfs}}, \\[5pt]
-{E_\gamma \over  \Delta E_{\,\mathrm{hfs}} } &\qquad E_\gamma \ll  \Delta E_{\,\mathrm{hfs}}.
 \end{array}\right.\nn
a_e(E_\gamma) &=& \left\{ \begin{array}{cl}
1 & \qquad E_\gamma \gg m \alpha^2, \\[5pt]
{ 2 E_\gamma \over m \alpha^2 } & \qquad E_\gamma \ll m \alpha^2.
 \end{array}\right.
\label{4.5}
\end{eqnarray}
In the limit $E_\gamma \gg m\alpha^2$ the Ore-Powell spectrum (Eq.~(\ref{2.4}))
is recovered, and it is shown that the sum 
of the magnetic and electric dipole transitions in the effective theory 
gives the full theory amplitude at leading order in the non-relativistic expansion.
The matching condition, which is the difference of the two results, vanishes~\cite{our}
and there is no additional three-photon annihilation term in the  NRQED Hamiltonian.

\begin{figure}[!t]
\begin{center}
\vspace{-1cm}
\hspace{-0.8cm}
\includegraphics[width=8.2cm]{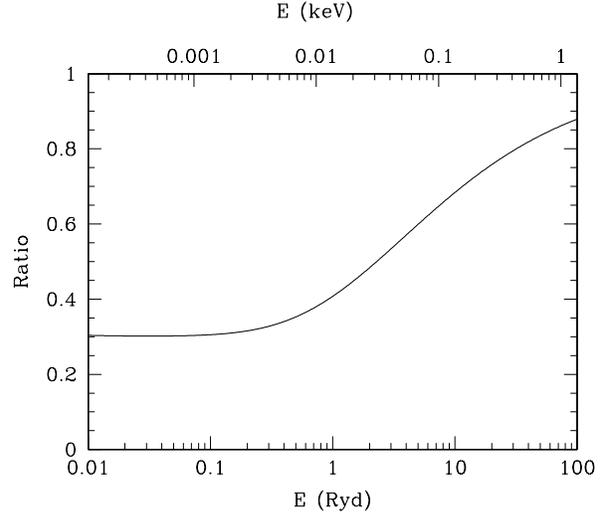}
\vspace{-1.2cm}
\caption{Ratio of the o-Ps decay spectrum including binding energy corrections to the
Ore-Powell spectrum (1 Ryd$=m\alpha^2/2$).
\label{fig:ratio}}
\end{center}
\vspace{-0.8cm}
\end{figure}

The ratio of the photon spectrum
to the Ore-Powell value is shown in Fig.~\ref{fig:ratio} up to $E_\gamma\sim m
\alpha/2$.
At energies large compared with the binding energy, $a_e$ and $a_m$ approach their values
in Eq.~(\ref{4.5}), and the spectrum reproduces the QED one. The magnetic and
electric dipole terms contribute in the ratio $3:7$. At energies small compared with the
binding energy, the electric dipole transitions
decouple, and one is left with the magnetic term 
due to the p-Ps resonance contribution, of $3/10$ of the Ore-Powell value.
At energies much smaller than the hyperfine splitting, the p-Ps state also decouples, and
the decay rate vanishes as $E_\gamma^3$, as predicted by Low's theorem.

The results are consistent with the Ore-Powell spectrum and with Low's theorem. They
include binding effects in a systematic expansion in powers of $v$. 

\vspace*{0.2cm}
\noindent
{\bf Acknowledgements}

We wish to thank S. Narison and his team 
for the organization of the QCD 04 conference.
This work has been supported in part by EU Contract No. 
HPRN-CT-2002-00311 (EURIDICE), by MCYT (Spain) under grant FPA2001-3031, and
by EU ERDF funds.

\end{document}